\documentclass{article}
\usepackage{graphicx}
\usepackage{adjustbox}      
\usepackage[numbers]{natbib}

\usepackage[final]{neurips_2023}

\usepackage[utf8]{inputenc} 
\usepackage[T1]{fontenc}    
\usepackage{hyperref}       
\usepackage{url}            
\usepackage{booktabs}       
\usepackage{amsfonts}       
\usepackage{nicefrac}       
\usepackage{microtype}      
\usepackage{xcolor}         

\title{Hierarchical Cross-entropy Loss for Classification of Astrophysical Transients}

\author{%
  V.~Ashley Villar \\
  Center for Astrophysics \textbar{} Harvard \& Smithsonian\\
  \texttt{ashleyvillar@cfa.harvard.edu} \\
\AND
Kaylee de Soto \\
Center for Astrophysics \textbar{} Harvard \& Smithsonian\\
\AND
Alex Gagliano \\
 The NSF AI Institute for Artificial Intelligence and Fundamental Interactions \\ 
}

\begin{document}

\maketitle

\begin{abstract}
Astrophysical transient phenomena are traditionally classified spectroscopically in a hierarchical taxonomy; however, this graph structure is currently not utilized in neural net-based photometric classifiers for time-domain astrophysics. Instead, independent classifiers are trained for different tiers of classified data, and events are excluded if they fall outside of these well-defined but flat classification schemes. Here, we introduce a weighted hierarchical cross-entropy objective function for classification of astrophysical transients. Our method allows users to directly build and use physics- or observationally-motivated tree-based taxonomies. Our weighted hierarchical cross-entropy loss directly uses this graph to accurately classify all targets into any node of the tree, re-weighting imbalanced classes. We test our novel loss on a set of variable stars and extragalactic transients from the Zwicky Transient Facility, showing that we can achieve similar performance to fine-tuned classifiers with the advantage of notably more flexibility in downstream classification tasks. 
\end{abstract}

\section{Introduction}

The Zwicky Transient Facility (ZTF; \cite{bellm2018zwicky}) currently discovers thousands of astronomical transients annually. By the end of 2025, the Vera C. Rubin Observatory will see first light and begin discovering \textit{millions} of transients annually. With exponentially increasing discovery rates of transient astrophysical phenomena in the night sky, the need for rapid and accurate classification continues to grow. These transient phenomena--supernovae, variable stars, active galactic nuclei, etc.--must be classified from sparse, multi-variate light curves (flux as a function of time and wavelength). Much work has been done to extract meaningful features from these phenomena for classification e.g., \cite{villar2020superraenn, boone2021parsnip, sanchez2021alert, gagliano2023first}. However, less focus has been given to the highly structured taxonomy of astrophysical transients. 

Transients, and indeed general astrophysical phenomena, are naturally ordered in hierarchies. For example, Type IIP supernovae (from red supergiant stars) belong to the parent class of hydrogen-rich supernovae, which themselves are a child of the core-collapse supernova class, which itself is a member of the supernova class. Similar hierarchies exist among variable stars. For example, a Be star is (typically) a main sequence star which itself is a type of variable star. Importantly, often events can only be classified into a general class (e.g., supernova-like) and lack a more specific label. Without a hierarchical structure built into the classification scheme to account for broad labels, large fractions of the training sets are often removed. To date, there has been no presentation for a hierarchical loss function to naturally account for this known tree-like taxonomy in classification. Here, we present such a loss.

Multi-tiered classifiers have been presented in the literature. The ALeRCE ZTF broker (a system for rapid classification of ZTF alerts) utilizes a set of two-tiered random forest classifiers which classify a combination of transient, stochastic and periodic phenomena \cite{sanchez2021alert}. The first random forest sorts objects into each of these categories. Then, at the second tier, these three categories then use separate random forest classifiers to further classify events into a (flat) set of labels. A similar setup is utilized in \cite{cheung2021new} to classify the ZTF catalog of periodic variables and by \cite{hosenie2019comparing} for variable stars in the CRTS survey. Both methods utilize re-sampling (e.g., SMOTE \cite{chawla2002smote}) to account for the heavily imbalanced training sets. Such methods may include unphysical realizations of transients, leading to worse performance at test time.

Here we present a new weighted hierarchical cross-entropy loss for classification. By explicitly encoding the natural graph taxonomy of our existing transient datasets, we are able to utilize the vast majority of our data in training and train only a single neural network for all classification.

\section{Weighted Hierarchical Cross Entropy}

First we introduce the weighted hierarchical cross-entropy loss (WHXE), which will be used in this work. The unweighted HXE was first presented in \cite{bertinetto2020making} and explored in the context of hierarchical image classification. Here, we modify this loss to include class weights, to account for the imbalanced classes often seen in astrophysical datasets.

To use the WHXE, we first design the graph associated with the astrophysical classification taxonomy. Specific examples will be shown in the following Section (for variable stars and supernovae). In general, the graph is required to be a directed tree (i.e., a directed, connected and acyclic graph).  

With a defined graphical structure, we next outline the WHXE objective function. To begin, we define the traditional categorical cross-entropy, typically used to optimize classifiers:
\begin{equation}
    \mathcal{L}= - \sum_i^C t_i \log(p(c_i)),
\end{equation}
where $C$ is the total number of SN classes which $c_i$ enumerates and $t_i$ is an indicator variable for the \textit{true} underlying transient class. Cross-entropy maximizes the total probability that the correct classes are returned by the classifier, with no penalty for how the probability is distributed over the incorrect classes.

 We denote the probability of a transient belonging to class $c$ as $P(c)$. We represent a class on a leaf node as $c^{(0)}$, while a class on the root (e.g., the ``Object'' parent class) is represented as $c^{(H)}$, with $H$ being the height of the tree. The probability for class $c$ can then be factorized as:
\begin{equation}
    p(c) = \prod^{H-1}_{h=0}p(c^{(h)}|c{(h+1)}).
\end{equation}
Note that the probability of the root class (the parent ``Object" label) is eliminated, as it assumed that each event is an astrophysical object with $P(c^{(H)})=1$. 

The conditional probabilities are therefore:
\begin{equation}
   p(c^{(h)}|c{(h+1)}) = \frac{\sum_{c_a\in\mathrm{Children}(c^{(h)})}p(c_a)}{\sum_{c_b\in\mathrm{Children}(c^{(h+1)})}p(c_b)},
\end{equation}
where $\mathrm{Children}(c)$ represents the set of child nodes of class $c$ (here indexed by $a$ and $b$). For example, the probability that a SN is H-rich given that it is a CCSN is equal to the probability that it is H-rich (vs H-poor) divided by the probability that it a CCSN (vs a Type Ia).

We now incorporate this information into the traditional definition for the categorical cross-entropy, by building a similar hierarchical loss with this factorized probability. The WHXE is defined as:
\begin{equation}
    \mathcal{L}_\mathrm{WHXE}(p,c)=-\sum_{h=0}^{H-1}W(c^{(h)})\lambda(c^{(h)})\log p(c^{(h)}|c^{(h+1)}),
\end{equation}
 where $W(c^{(h)})$ reweights each object by its class fraction. Here, we choose to reweight each class by $N_\mathrm{All}/(N_\mathrm{Labels}\times N_\mathrm{c})$, where $N_\mathrm{All}$ is the total number of events in the dataset, $N_\mathrm{Labels}$ is the number of unique classes, and $N_\mathrm{c}$ is the number of events of class $c$. Note that this naturally means that parent classes have a lower weighting than leaf nodes (as parent classes include all sibling leaves). 
 
 Furthermore, $\lambda(c^{(h)})$ is a second weighting which emphasizes different levels of the tree. \cite{bertinetto2020making} suggest the following form for $\lambda(c^{(h)})$:
\begin{equation}
    \lambda(c) = \exp(-\alpha h(c)),
\end{equation}
where $\alpha$ is a free parameter. Larger values of $\alpha$ weight the top of the hierarchy more strongly (i.e., Ia vs core-collapse supernova classification); lower values of $\alpha$ weight each level of the hierarchy equally, emphasizing fine-grained classifications. $\alpha$ is therefore a hyperparameter of our model which we can optimize; here, we fix $\alpha=0.5$, which somewhat favors parent classification over fine-grained classification. We note that a value of $\alpha=0$ would treat each node equally, while a values of $\alpha>1$ greatly favors parent classification. 

Our classifier must output a vector with a length equal to the number of vertices in the graph. A softmax function is then applied to each set of siblings such that the neural network output can be interpreted as a conditional pseudo-probability. 

We provide a simple Python script which implements the WHXE in Pytorch, showcasing the experiments outlined below\footnote{\url{https://github.com/villrv/hxe-for-tda/}}. We provide a number of helper functions that allow users to, after generating a graph, easily implement the WHXE and calculate probabilities for any node of the graph using the trained model. 

\section{Experiments \& Results}

\subsection{Datasets}

We use two datasets to showcase the utility of WHXE. All data and code associated with this project are available on GitHub.

\begin{figure}
  \centering
\includegraphics[width=\linewidth]{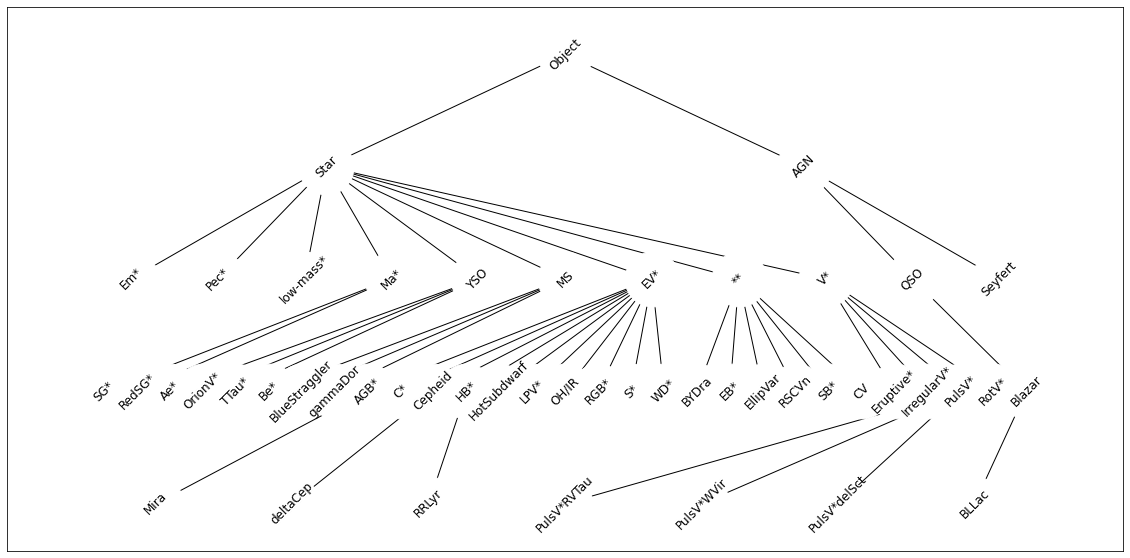}\label{fig:varg}
  \caption{Graphical representation of ZTF variable star dataset. Note here that ``*'' represents ``star"; i.e., ``**" should be read as ``binary star", and ``C*" is ``carbon star."}
\end{figure}

\textbf{Variable Stars:} We use the ZTF periodic variable star catalog, first presented in \cite{chen2020zwicky}. After discovering that this catalog contains non-stellar objects (e.g., active galactic nuclei), \cite{cheung2021new} produced a new set of labels, which we use in this work. This set includes 46,078 objects which were successfully cross-matched against the SIMBAD astronomical database. The graph structure of this dataset is shown in Figure~\ref{fig:varg}. We remove some objects which do not fall into an obvious time-variable  category (e.g., galaxy, unknown types) and limit our final dataset size to 45,748 objects. We map some objects into standard classes (e.g., ``candidate eclipsing binary" stars are relabeled as ``eclipsing binary" stars). We use the 19 features presented in \cite{cheung2021new}, which include standard measurements such as variability period, amplitude mean magnitudes, etc. The dataset is imbalanced, with the majority class (eclipsing binary stars) making up $\sim40$\% of the dataset, and the minority classes containing just one member each (blazars and supergiant stars). 

\begin{figure}
  \centering
\includegraphics[width=\linewidth]{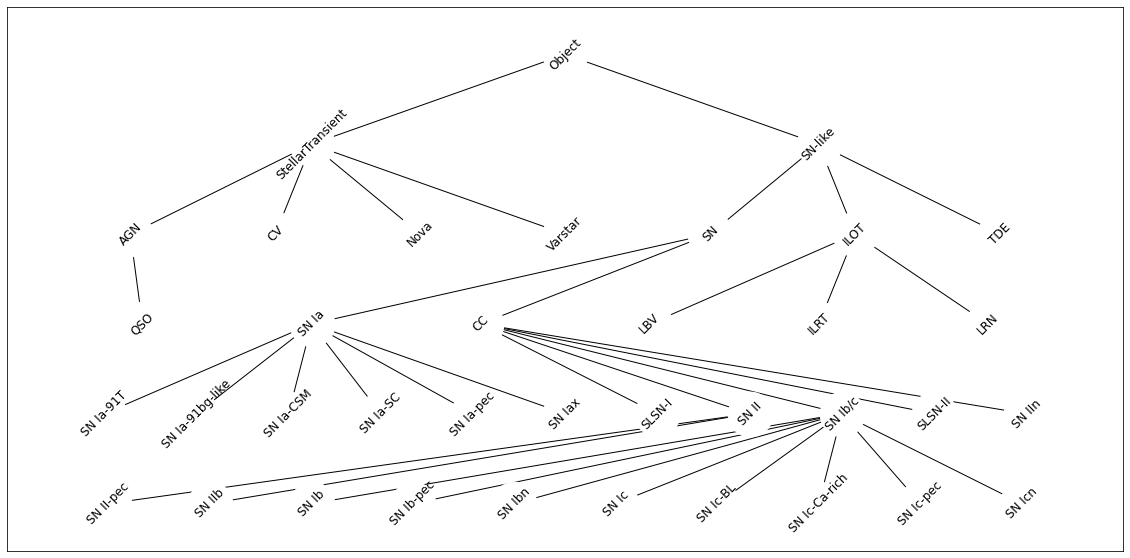}\label{fig:sng}
  \caption{Graphical representation of ZTF supernova dataset.}
\end{figure}

\textbf{Supernovae:} We additionally use a dataset of supernovae observed with ZTF \cite{desoto}. This set contains 4,214 events with transient spectroscopic labels provided by the Transient Name Server. The graph representation of this data is shown in Figure~\ref{fig:sng}. We remove Type I objects from this dataset; the Type I classification indicates a lack of hydrogen in the spectrum but does not point to a specific progenitor channel (such a type can live on multiple trees of our graph, see \cite{filippenko1997optical} for a summary of supernova spectroscopy classification). Due to this ambiguity, we remove these objects. We additionally remove some stellar flares and non-transients. Finally, we map some objects to standardized classes (e.g., SN IIP/L to SN II). Our final dataset consists of 4,206 events. The features consists of fitted light curve parameters to both the $g$- and $r$-bands light curves from ZTF, for a total of 15 features (see \cite{desoto} or \cite{hosseinzadeh2020photometric} for details on the model parameters). Again, our dataset is heavily imbalanced. Our majority class (Type Ia supernovae) makes up $\simeq65$\% of the dataset, while our minority classes (luminous red novae, peculiar Type Ic supernovae and intermediate luminosity transients) each contain only one sample.

\subsection{Experiments}

We train a multi-layer perceptron (MLP) classifier utilizing our WHXE for the two datasets outlined above. We train \textit{one} classifier per dataset for all experiments. For the variable star dataset, we train a MLP with two hidden layers of 20 neurons for 500 epochs with a batch size of 4096 and learning rate of 0.01. For the supernova dataset, we use two hidden layers with 15 neurons each, training with the AdamW optimizer \cite{kingma2014adam} for 1000 epochs with a batch size of 128, learning rate of 0.001 and weight decay of 0.001.  We split our datasets into training/test sets with a 66/33 split.

In both of our experiments, we compare to baseline MLP classifiers trained specifically \textit{for each task}, using the appropriate subset of data. We track the macro/micro F1 score (the geometric mean of the precision and recall) and fraction of sample used in training. Our results are summarized in Table~\ref{tab:tab}. We report uncertainties as the standard deviation of these scores using three random seeds for the train/test split and training pipeline.

\textbf{Multi-class tests:} First, we compare the performance of our WHXE models to the fine-tuned models on the (primarily) leaf nodes of the graph. For variable stars, we include all possible leaf nodes (a total of 35 outputs). Given the small number of members for some classes, it is possible in this experiment that some objects are entirely excluded from the training/test sets. We find statistically comparable performance between the baseline and multi-class model, both achieving micro-averaged F1 scores of $\simeq0.9$; both have notably lower macro-averaged F1 scores ($\simeq0.25$), significantly reduced by the minority classes with few samples. However, we note that the baseline model required us to remove $\simeq66$\% of the dataset to train due to missing relevant labels on many objects.

For the supernova dataset, rather than classify on the leaves, we attempt classification on the standard 5-class setup: SN-Ibc, SN-IIP/L, SN-IIn, SLSN-I and SN-Ia. Again, the WHXE and baseline methods perform similarly well, achieving a macro-averaged F1 score of $\simeq0.45$ and a micro-averaged F1 score of $\simeq0.84$. It is worth noting, however, that in training the baseline, 45\% of the dataset was removed.

\textbf{Top-level- tests:} We next compare the performance of our WHXE models to the fine-tuned models trained on the top ``parent" classes of the graph. In this test, all objects must fall into one of these two parent classes. For variable stars, this includes active galactic nuclei and stars. This top class if notably imbalanced, with active galactic nuclei making up just $\simeq1$\% of the sample. We find that, again the WHXE and baseline methods perform similarly, achieving macro-averaged F1 scores of $\simeq0.95$.

For the supernova dataset, there are three top-level categories to which all objects belong: "Stellar Transient" and "Supernova-like (SN-like) Transient". Here, only $\simeq3$\% of our sample fall under the stellar transient class. Again, our WHXE and baseline models achieve similar F1 scores within uncertainties (with the macro-averaged F1 score being $\simeq0.75$. Our WHXE model achieves an average F1 score $\simeq0.53$, while our fine-tuned baseline model achieves $\simeq0.49$.

\begin{table}
  \caption{Experimental Results. Parenthetical numbers are uncertainties on the final digit(s). }
  \label{tab:tab}
  \centering
\begin{adjustbox}{center}

  \begin{tabular}{llllllll}
    \toprule
    Task     & & & Variable Dataset   &  & &Supernova Dataset &  \\
    \cmidrule(r){3-5} \cmidrule(r){6-8} 

    &    & F1 (macro) & F1 (micro)& frac. & F1 (macro)& F1 (micro) & frac. \\

    \midrule
    Multi-class & WHXE & 0.232 (4)  & 0.894 (2) & 1.0 &  \textbf{0.485 (4)}& \textbf{0.855 (1)}& 1.0 \\
     & Baseline& \textbf{0.256 (2)}  & \textbf{0.899 (1)} &0.33&  0.447 (14) &  0.835 (8)& 0.45 \\
    Top-level    & WHXE& \textbf{0.966 (2)}  &  \textbf{0.998 (1)} & 1.0 & \textbf{0.753 (28)}& \textbf{0.976 (3)}& 1.0\\
     & Baseline& \textbf{0.966 (6)} & \textbf{0.998 (1)} & 1.0 & 0.716 (45) & 0.972 (4)& 1.0\\
    \bottomrule
  \end{tabular}
\end{adjustbox}

\end{table}

\section{Conclusions}
Building on the HXE presented in \cite{bertinetto2020making}, we have introduced a new loss function, the weighted hierarchical cross-entropy, for classification of astrophysical objects with naturally hierarchical taxonomies. We have implemented the WHXE as a loss function for a simple MLP-based classifier for a set of ZTF variable stars and supernova-like transients. We find that, compared to a fine-tuned baseline, we are able to achieve similar results for a number of specialized classification tasks. Importantly, our new objective function allows users to utilize all data in training and produce class pseudo-probabilities for any node of the classification graph. With the upcoming Vera C. Rubin Observatory, efficient classification methodologies which account for natural structure within astronomical datasets will be essential as the training of fine-tuned classifiers becomes increasingly computationally expensive. In future work, we will explore the ability of our method to efficiently isolate rare transients, and to understand its performance on a joint variable star/extragalactic transient dataset. 

\acksection
We thank anonymous reviewers for helpful feedback which improved this manuscript. VAV and KdS acknowledge support by the NSF through grant AST-2108676. This work is supported by the National Science Foundation under Cooperative Agreement PHY-2019786 (The NSF AI Institute for Artificial Intelligence and Fundamental Interactions, http://iaifi.org/).

{
\small
\bibliographystyle{plain}
\bibliography{main}

\begin{thebibliography}{10}

\bibitem{bellm2018zwicky}
Eric~C Bellm, Shrinivas~R Kulkarni, Matthew~J Graham, Richard Dekany, Roger~M Smith, Reed Riddle, Frank~J Masci, George Helou, Thomas~A Prince, Scott~M Adams, et~al.
\newblock The zwicky transient facility: system overview, performance, and first results.
\newblock {\em Publications of the Astronomical Society of the Pacific}, 131(995):018002, 2018.

\bibitem{bertinetto2020making}
Luca Bertinetto, Romain Mueller, Konstantinos Tertikas, Sina Samangooei, and Nicholas~A Lord.
\newblock Making better mistakes: Leveraging class hierarchies with deep networks.
\newblock In {\em Proceedings of the IEEE/CVF conference on computer vision and pattern recognition}, pages 12506--12515, 2020.

\bibitem{boone2021parsnip}
Kyle Boone.
\newblock Parsnip: Generative models of transient light curves with physics-enabled deep learning.
\newblock {\em The Astronomical Journal}, 162(6):275, 2021.

\bibitem{chawla2002smote}
Nitesh~V Chawla, Kevin~W Bowyer, Lawrence~O Hall, and W~Philip Kegelmeyer.
\newblock Smote: synthetic minority over-sampling technique.
\newblock {\em Journal of artificial intelligence research}, 16:321--357, 2002.

\bibitem{chen2020zwicky}
Xiaodian Chen, Shu Wang, Licai Deng, Richard De~Grijs, Ming Yang, and Hao Tian.
\newblock The zwicky transient facility catalog of periodic variable stars.
\newblock {\em The Astrophysical Journal Supplement Series}, 249(1):18, 2020.

\bibitem{cheung2021new}
Siu-Hei Cheung, V~Ashley Villar, Ho-Sang Chan, and Shirley Ho.
\newblock A new classification model for the ztf catalog of periodic variable stars.
\newblock {\em Research Notes of the AAS}, 5(12):282, 2021.

\bibitem{desoto}
Kaylee de~Soto and V.~Ashley Villar.
\newblock Superphot+: Real-time fitting and classification of ztf supernova light curves.
\newblock {\em in prep.}, 2023.

\bibitem{filippenko1997optical}
Alexei~V Filippenko.
\newblock Optical spectra of supernovae.
\newblock {\em Annual Review of Astronomy and Astrophysics}, 35(1):309--355, 1997.

\bibitem{gagliano2023first}
Alexander Gagliano, Gabriella Contardo, Daniel~Foreman Mackey, Alex~I Malz, and Patrick~D Aleo.
\newblock First impressions: Early-time classification of supernovae using host galaxy information and shallow learning.
\newblock {\em arXiv preprint arXiv:2305.08894}, 2023.

\bibitem{hosenie2019comparing}
Zafiirah Hosenie, Robert~J Lyon, Benjamin~W Stappers, and Arrykrishna Mootoovaloo.
\newblock Comparing multiclass, binary, and hierarchical machine learning classification schemes for variable stars.
\newblock {\em Monthly Notices of the Royal Astronomical Society}, 488(4):4858--4872, 2019.

\bibitem{hosseinzadeh2020photometric}
Griffin Hosseinzadeh, Frederick Dauphin, V~Ashley Villar, Edo Berger, David~O Jones, Peter Challis, Ryan Chornock, Maria~R Drout, Ryan~J Foley, Robert~P Kirshner, et~al.
\newblock Photometric classification of 2315 pan-starrs1 supernovae with superphot.
\newblock {\em The Astrophysical Journal}, 905(2):93, 2020.

\bibitem{kingma2014adam}
Diederik~P Kingma and Jimmy Ba.
\newblock Adam: A method for stochastic optimization.
\newblock {\em arXiv preprint arXiv:1412.6980}, 2014.

\bibitem{sanchez2021alert}
P~S{\'a}nchez-S{\'a}ez, I~Reyes, C~Valenzuela, F~F{\"o}rster, S~Eyheramendy, F~Elorrieta, FE~Bauer, G~Cabrera-Vives, PA~Est{\'e}vez, M~Catelan, et~al.
\newblock Alert classification for the alerce broker system: The light curve classifier.
\newblock {\em The Astronomical Journal}, 161(3):141, 2021.

\bibitem{villar2020superraenn}
V~Ashley Villar, Griffin Hosseinzadeh, Edo Berger, Michelle Ntampaka, David~O Jones, Peter Challis, Ryan Chornock, Maria~R Drout, Ryan~J Foley, Robert~P Kirshner, et~al.
\newblock Superraenn: A semisupervised supernova photometric classification pipeline trained on pan-starrs1 medium-deep survey supernovae.
\newblock {\em The Astrophysical Journal}, 905(2):94, 2020.

\end{thebibliography}
}


\end{document}